\documentclass[10pt,conference]{IEEEtran}
\IEEEoverridecommandlockouts
\usepackage{graphicx}
\usepackage{url}
\usepackage{amsmath,amssymb,amsfonts}
\usepackage{booktabs}
\usepackage{listings}
\usepackage{multirow}
\usepackage{braket}
\usepackage{ascmac}
\usepackage[T1]{fontenc}

\def\BibTeX{{\rm B\kern-.05em{\sc i\kern-.025em b}\kern-.08em
    T\kern-.1667em\lower.7ex\hbox{E}\kern-.125emX}}

\newcommand{\conclusionbox}[1]{\begin{center}\fbox{\begin{minipage}{0.96\linewidth}\small Summary: #1\end{minipage}}\end{center}}

\begin{document}
\title{Database Reordering for Compact Grover Oracles with ESOP Minimization}

\author{\IEEEauthorblockN{
    Yusuke Kimura\IEEEauthorrefmark{1} and Yutaka Takita\IEEEauthorrefmark{1}}
    \IEEEauthorblockA{\textit{Fujitsu Limited, Japan}\IEEEauthorrefmark{1}}
    \{yusuke-kimura, takita.yutaka\}@fujitsu.com\IEEEauthorrefmark{1}
}
\maketitle

\begin{abstract}
QALG, RESP paper

Grover's algorithm searches for data satisfying a desired condition in an unstructured database. This algorithm can search a space of size $N$ in $\sqrt{N}$ queries, thereby achieving a quadratic speedup. However, within the Grover oracle circuit that is repeatedly applied, the quantum state preparation circuit---which embeds database information into quantum states---suffers from a large gate count and circuit depth. To address this problem, we propose reducing the quantum state preparation circuit by reordering the database. Specifically, we consider a Quantum Read-Only Memory (QROM), where data are assigned to addresses, and assume that the address assignment of data can be freely permuted. By applying Exclusive Sum-of-Products (ESOP) minimization to the resulting truth table, we reduce the quantum circuit. Although the resulting circuit logic differs from the original, the state preparation remains correct in the sense that every desired datum is encoded at some address. Furthermore, we propose a proxy metric that estimates circuit size without compilation, and combine it with simulated annealing to efficiently find a near-optimal data ordering.

In our experiments, an exhaustive search over all orderings for databases of size $N=8$ reveals that circuit size varies by up to approximately a factor of two depending on the ordering, demonstrating the utility of reordering. Compared with applying ESOP minimization without reordering, simulated annealing reduces the circuit size by approximately 30\% and yields circuits close to optimal. For $N=64$ and $128$, simulated annealing is shown to discover smaller circuits compared with random search.
\end{abstract}

\section{Introduction}

Quantum computers can achieve computational advantages over classical computers for certain problems by exploiting quantum superposition and interference. As a representative algorithm, Grover search provides a quadratic speedup for unstructured search spaces~\cite{grover1996}. However, the practical bottleneck often lies not in the number of iterations itself but in the resource cost of the oracle circuit that is repeatedly applied in each iteration~\cite{10.1145/3571725,PhysRevX.14.041029}. In particular, in fault-tolerant settings, non-Clifford gates (typically $T$ gates) are expensive, and reducing $T$-count and $T$-depth is of critical importance~\cite{fowler2012surfacecode}.

When applying Grover search to database search problems, it is necessary to look up data corresponding to a given address or to embed data as quantum states. Such operations can be described as QROM (Quantum Read-Only Memory) circuits~\cite{qrom}, and the cost of these circuits often dominates the overall resource budget. \cite{phalak2022iccd} reported circuit optimization techniques for QROM. QROM circuits contain many multi-controlled X (MCX) gates, and decomposition techniques for multi-controlled operations~\cite{barenco1995} are used to realize them on hardware. Moreover, the workflow of converting Boolean functions into an ESOP (Exclusive Sum of Products) representation, minimizing the representation, and implementing it as a Toffoli cascade is well established in classical circuit optimization~\cite{8758744}, with EXORCISM being a representative ESOP minimization method~\cite{mishchenko2001}.

On the other hand, most existing studies optimize the implementation of quantum circuits under the assumption that the given truth table---i.e., the mapping between addresses and data---is fixed. This paper focuses on the observation that, in Grover search, the ordering of the database is often meaningless. That is, if the search targets are given as a set and the measured address can be classically mapped back to the original identifier, the mapping between addresses and data can be freely permuted in advance. This degree of freedom implies that the function represented by the oracle can be reshaped, potentially leading to significantly different results from ESOP minimization. In other words, we can add a new axis---data ordering (address assignment)---to the search space of circuit optimization. 

In this paper, we formalize this idea as Database Reordering and propose an integrated optimization method for QROM-style data-encoding circuits that combines (i) database reordering (address relabeling), (ii) ESOP minimization, (iii) evaluation via a proxy metric that avoids compilation, and (iv) permutation search using simulated annealing (SA). Since the permutation search space is $N!$ and exhaustive search is impractical, we employ SA~\cite{kirkpatrick1983} to obtain good orderings within a limited evaluation budget. Moreover, because running quantum circuit compilation at every step of the search loop is computationally prohibitive, we introduce a proxy metric computed from the statistics of the ESOP representation after minimization, thereby accelerating the search. Our method does not replace existing QROM optimization or Grover oracle optimization but can be incorporated as a preceding stage. It is therefore compatible with the aforementioned existing techniques.

The contributions of this paper are as follows.
\begin{itemize}
\item We observe that, in unstructured search, the ordering of data can be freely changed, and introduce data ordering (address assignment) as an optimization axis for oracle reduction.
\item We define a proxy metric that can predict $T$-gate count without compilation, making the search feasible within practical time.
\item We demonstrate that SA-based search consistently discovers lower-cost orderings than random search.
\item Through experiments, we confirm that for $N=8$, circuit size (as measured by the proxy) varies by up to approximately a factor of two depending on the ordering, that the proposed method yields smaller circuits than ESOP minimization with a fixed ordering, and that the advantage is maintained for larger $N$.
\end{itemize}

The remainder of this paper is organized as follows. Sec.~\ref{sec:background} reviews the relationships among Grover search, QROM, ESOP minimization, and quantum circuit optimization. Sec.~\ref{sec:motivation} describes the motivation leading to our proposed method. Sec.~\ref{sec:proposed} presents the proposed method, including the problem formulation of Database Reordering, the proxy metric, and SA-based search. Sec.~\ref{sec:experiments} describes the experimental setup and results. Sec.~\ref{sec:limitations} discusses the limitations of the proxy metric, search budgets, and other considerations. Finally, Sec.~\ref{sec:conclusion} concludes the paper and outlines future work.

\section{Background}\label{sec:background}
\subsection{Grover Search and Oracles}
Grover search is a quantum algorithm for finding elements that satisfy a given condition within an unstructured set of candidates~\cite{grover1996}. By alternately applying an oracle, which determines whether each candidate is a solution, and a diffusion operator, which amplifies the probability of the solution state. Amplitude amplification generalizes this idea, providing the foundation for understanding Grover search in the broader context of success probability amplification~\cite{brassard2002}.

The advantage of Grover search is often explained in terms of a quadratic reduction in the number of queries. However, when actually implementing the algorithm as a quantum circuit, the size of the oracle applied at each iteration tends to become the bottleneck~\cite{10.1145/3571725,PhysRevX.14.041029}.

Furthermore, in fault-tolerant quantum computation, the implementation cost of non-Clifford gates is high, and $T$-count and $T$-depth serve as the primary evaluation metrics~\cite{fowler2012surfacecode}. Consequently, reducing the oracle circuit is important not merely for gate count reduction but also for resource reduction in fault-tolerant quantum computation. Decomposition methods for multi-controlled operations and Toffoli-type gates have been studied extensively~\cite{barenco1995,jones2013toffoli}, and these results form the basis for oracle optimization.

\subsection{QROM and Data Encoding Circuits}
The data encoding circuit considered in this paper can be expressed as a QROM (Quantum Read-Only Memory)~\cite{qrom}. A QROM is a circuit that writes fixed data to an output register based on the value specified by an address register. Denoting the address as $a$ and the fixed database as $D[\cdot]$, its basic operation is written as
\begin{equation}
\ket{a}\ket{0} \rightarrow \ket{a}\ket{D[a]}
\label{eq:bg_qrom}
\end{equation}
A crucial point is that $D[\cdot]$ is constant data known at compile time, not memory that changes at runtime. Therefore, a QROM can be treated as a circuit synthesis problem: how to embed a fixed database as a logic circuit.

This type of circuit is related to research on quantum memory in general. For example, \cite{giovannetti2008qram} proposed the QRAM architecture, which enables memory access with addresses in quantum superposition. In contrast, the QROM considered in this paper is a reference circuit for known data, not a dynamic quantum memory mechanism. Although the two differ in purpose and implementation assumptions, they share the common background that data loading in quantum computation is expensive.

The size of a QROM circuit grows rapidly as the address width and data width increase. For this reason, research has been conducted on reducing gate count and depth by devising the control structure and decomposition of QROM~\cite{phalak2022iccd}. Such research aims to implement a given database more efficiently and constitutes an important practical foundation for quantum search and quantum state preparation.

A QROM can also be viewed as a multi-output Boolean function. If the data width is $d$ bits, each output bit corresponds to a single Boolean function with the address as input. Thus, QROM synthesis can be reformulated as the problem of implementing multiple Boolean functions as a reversible circuit, and the ESOP minimization methods described below become applicable.

\subsection{ESOP Minimization and Reversible Circuit Synthesis}
ESOP (Exclusive Sum of Products) is a representation that expresses a Boolean function as an XOR sum of product terms, and it has been widely used in logic synthesis~\cite{mishchenko2001}. Each product term can be understood as a condition that specifies 0 or 1 for certain input variables and treats the rest as don't-cares. The advantage of the ESOP representation is that each product term directly corresponds to a ``conditionally flip the output'' operation, making the correspondence with quantum circuits straightforward~\cite{8758744}.

In the context of reversible and quantum circuits, each ESOP term can be implemented as an MCX or Toffoli-type gate. Therefore, the number of terms in an ESOP and the number of specified literals in each term are closely related to the final circuit size. The problem of how to decompose multi-controlled gates has itself been studied~\cite{barenco1995,jones2013toffoli}, and these results are important for understanding how the complexity of an ESOP representation affects quantum circuit cost.

ESOP minimization is generally a hard problem, and heuristic methods are commonly used in practice. EXORCISM is a representative example, known as an algorithm that can reduce ESOP representations relatively quickly~\cite{mishchenko2001}. Various methods for converting a given specification into a Toffoli cascade or a reversible gate network have also been proposed~\cite{miller2003,shende2003}, and these serve as the foundational techniques for automatic generation and optimization of quantum oracle circuits.

Optimization using ordering and permutation, which is the focus of this paper, has also been considered previously. Wille et al.~\cite{wille2009swop} showed that incorporating output permutation can reduce circuit size, and proposed Synthesis with Output Permutation (SWOP). Datta et al.~\cite{datta2014perm} reported methods that treat input and output orderings as search variables and use evolutionary computation or SA to reduce synthesis cost. These studies demonstrate the importance of ordering optimization as a concept; however, the idea of reordering the database itself, as proposed in this paper, is distinct.

\section{Motivation}\label{sec:motivation}

\begin{figure*}[t]
  \centering
  \includegraphics[width=0.9\linewidth]{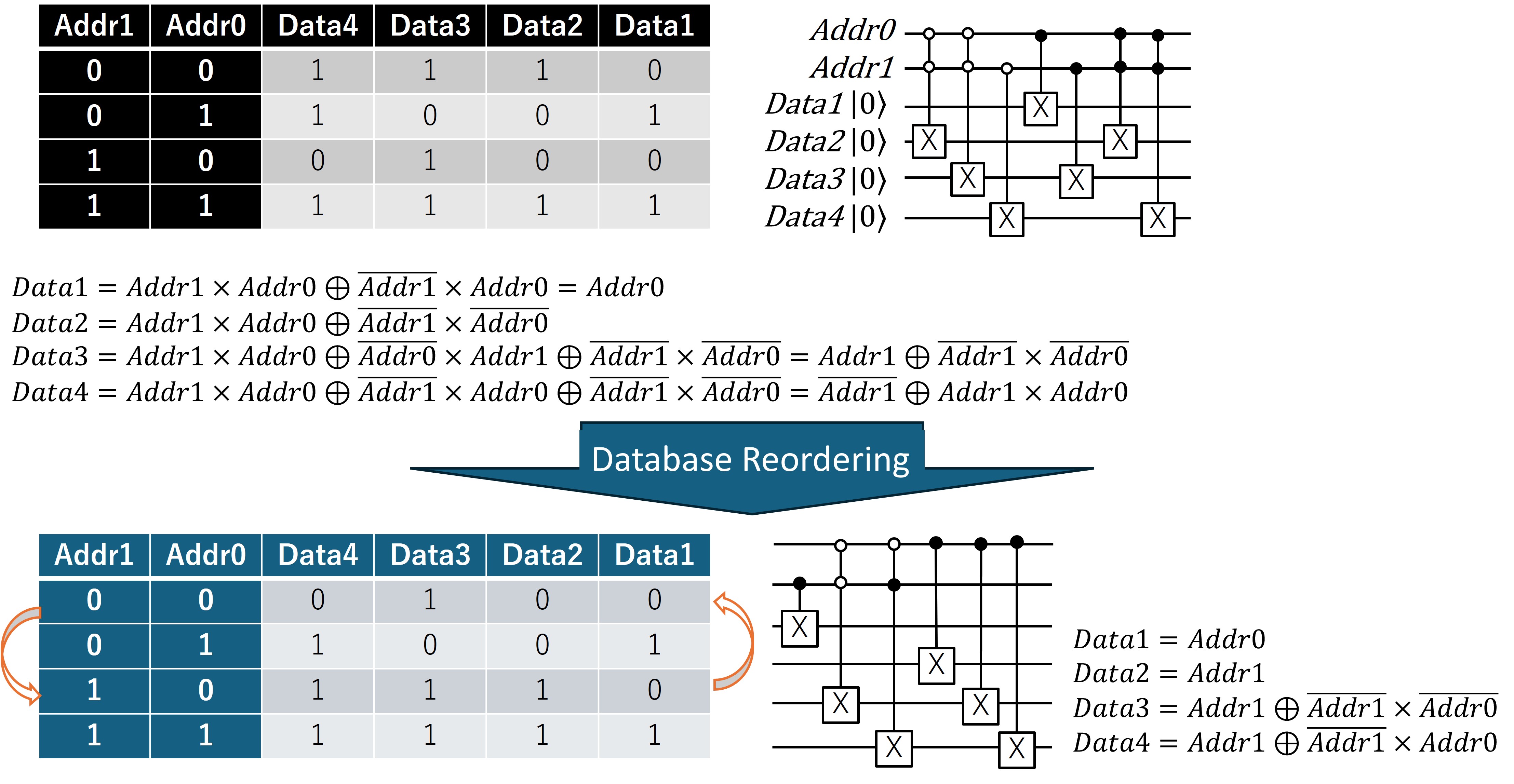}
  \caption{Motivation: Database reordering}
  \label{fig:motivation}
\end{figure*}

This work focuses on the observation that, even when the set of data to be embedded is the same, the circuit size can vary depending on which addresses the data are assigned to. In particular, in the ESOP minimization used in this paper, the positions of 1s in the truth table affect how product terms can be grouped. Consequently, even for the same data set, different address assignments can yield different circuits.

To illustrate this point intuitively, consider a simple example in which four 4-bit data items are embedded with 2 address-bits. The original database is given as in the upper left of Fig.~\ref{fig:motivation}. Here, data items 1110, 1001, 0100, and 1111 are assigned to addresses $00, 01, 10, 11$, respectively. By extracting the Boolean function for each output bit of this database, performing ESOP minimization, and converting the result to a quantum circuit, we obtain the circuit shown in the upper right.

A key observation is that, in many settings of Grover search and QROM-type data encoding, it suffices for all data to be correctly embedded at some address; the specific address to which each datum is assigned has no meaning. In such situations, one can change only the mapping between addresses and data while keeping the data set fixed. The table in the lower left of Fig.~\ref{fig:motivation} shows an example in which the data originally assigned to addresses $00$ and $10$ have been swapped. The data themselves are entirely unchanged; only the address assignment differs.

Applying the same ESOP minimization to this reordered database yields a smaller circuit. In this example, the original ordering required
\begin{equation}
4~\mathrm{CCX} + 3~\mathrm{CX} = 7~\mathrm{gates}
\label{eq:motivation_orig_cost}
\end{equation}
whereas the reordered version requires
\begin{equation}
2~\mathrm{CCX} + 4~\mathrm{CX} = 6~\mathrm{gates}
\label{eq:motivation_reordered_cost}
\end{equation}
reducing the total gate count by one. More importantly, the number of CCX gates---which tend to incur higher cost---decreases from four to two. In general, multi-controlled gates strongly affect the $T$-count and depth after decomposition, so this type of reduction is significant from an implementation cost perspective.

This example demonstrates that the result of ESOP minimization is determined not only by the set of data to be embedded but also by where those data are placed in the truth table. In other words, the database ordering itself, which has conventionally been regarded as fixed, can serve as a target for circuit optimization.

\section{Proposed Methods}\label{sec:proposed}
\subsection{Problem Setting: Database Reordering}
In this paper, we consider a fixed database with an address width of $n$ bits ($N=2^n$) and a data width of $d$ bits:
\begin{equation}
D = \{D[0], D[1], \ldots, D[N-1]\}, \quad D[a] \in \{0,1\}^{d},
\label{eq:method_database}
\end{equation}
The corresponding encoding circuit $U_D$ is expressed as
\begin{equation}
U_D \ket{a}\ket{0} = \ket{a}\ket{D[a]}
\label{eq:method_qrom}
\end{equation}
In Grover search, reducing this encoding circuit leads to a reduction in the overall oracle circuit size.

This paper assumes that, in unstructured database search where Grover's algorithm is employed, the ordering within the database is meaningless. For example, when searching for genomic sequences similar to a particular query sequence within a database, the storage order of records typically does not affect the search result. If the search targets are given as a set and the measured address can be classically mapped back to the original identifier after measurement, the mapping between addresses and data can be freely permuted in advance. We represent this permutation as a permutation $\pi$ on the address set $\{0,\ldots,N-1\}$. The reordered database $D_{\pi}$ is defined as
\begin{equation}
D_{\pi}[\pi(a)] = D[a]
\label{eq:method_reordered_db}
\end{equation}
In other words, the data originally stored at address $a$ is moved to the new address $\pi(a)$. When a new address $a'$ is obtained as a search result, the original address can be recovered as $\pi^{-1}(a')$.

The objective of this paper is to reduce the implementation cost of the encoding circuit by optimizing the database ordering. Ideally, we seek the permutation that minimizes the compiled circuit cost $C(\pi)$:
\begin{equation}
\pi^{\ast} = \arg\min_{\pi \in S_N} C(\pi)
\label{eq:method_true_objective}
\end{equation}
where $S_N$ denotes the set of all permutations of $N$ data items and has size $N!$. Therefore, exhaustive search is impractical for general $N$, and compiling to a quantum circuit and evaluating the $T$-count or depth for each candidate is also expensive. In Sec.~\ref{sec:proposed:proxy}, we introduce a fast proxy metric computable from the ESOP representation, and the permutation search is conducted using this proxy as the objective function. The search method is described in Sec.~\ref{sec:proposed:anneal}.

\subsection{Proxy Metric for Faster Search}\label{sec:proposed:proxy}
In the proposed method, many candidate orderings must be evaluated. However, generating a quantum circuit for each candidate and further compiling it into a Clifford+$T$ circuit to measure $T$-count or depth is computationally too expensive. Therefore, an evaluation metric that can rapidly approximate the compiled circuit size is needed.

In this paper, we use a proxy metric based on the ESOP representation. As described in Sec.~\ref{sec:proposed}, each output bit of a QROM can be regarded as a single Boolean function, so ESOP minimization can be performed for each output bit. An ESOP consists of multiple cubes (product terms), each representing the condition under which the output is flipped, specified by which input bits must be 0 or 1.

Please note that each cube corresponds to a single MCX gate in the quantum circuit. The more input bits that are specified in a cube, the larger the number of controls on the corresponding MCX gate. In general, gates with more controls incur higher decomposition costs. We therefore propose counting the number of specified literals in each cube as a simple way to measure ESOP complexity.

For example, the 3-input cube $01{-}$ indicates that the first input must be 0, the second input must be 1, and the third is don't-care. In this case, two inputs are specified, so the complexity of this cube is counted as 2. Conversely, a cube such as $1{-}{-}$, where only one input is specified, can be implemented with simpler control.

Based on the above, we define the proxy score $S(\pi)$ for an ordering $\pi$ as the total number of specified literals across all cubes obtained after ESOP minimization. Letting $\mathcal{C}_j(\pi)$ denote the set of cubes for output bit $j$ and $\mathrm{lit}(c)$ denote the number of specified literals of cube $c$, the proxy is given by
\begin{equation}
S(\pi) = \sum_{j=1}^{d}\sum_{c \in \mathcal{C}_j(\pi)} \mathrm{lit}(c)
\label{eq:proxy_simple}
\end{equation}

This metric intuitively reflects both the number of cubes and the complexity of each cube simultaneously. If the number of cubes is large, the sum increases; even when the number of cubes is the same, a larger proportion of complex cubes also increases the value. Therefore, a smaller $S(\pi)$ generally indicates that the corresponding ESOP is easier to implement.

There are two reasons for using this proxy in this paper. First, it can be computed immediately from the output of ESOP minimization, making evaluation during the search very lightweight. Second, because each cube corresponds to an MCX-type gate, the proxy is expected to correlate to some extent with the actual circuit resource cost, particularly $T$-count.
For example, when compiling MCX gates with Qiskit, \cite{Khattar_2025} is used by default. In this method, an $N$-controlled MCX gate is decomposed into $8n-12$ $T$ gates.
Of course, the final $T$-count also depends on the MCX decomposition method and compiler optimizations, so the proxy is not an exact measure of resource cost. Nevertheless, we consider it useful as a simple metric for comparing many candidate orderings.

When using the proposed method, this proxy is used to search for a good ordering, and only the final selected candidates are compiled. In Sec.~\ref{sec:experiments}, we also verify the correlation between this proxy and the compiled $T$-count.

Note that if one wishes to reflect finer-grained hardware costs, a different metric can be used. In addition, cubes with a specified literal count of 1 correspond to CX gates, which can be implemented as Clifford gates and do not require $T$ gates. Therefore, the proxy may overestimate the $T$-count in some cases.

\subsection{Optimal Order Search with Simulated Annealing (SA)}\label{sec:proposed:anneal}
The proxy defined in the previous section is a metric for rapidly computing an approximate circuit size for a given ordering. However, the number of orderings to be evaluated is very large. For a database of size $N$, the total number of candidate orderings is $N!$, so examining all of them is impractical except for small-$N$ cases. Therefore, in this paper, we employ SA as a search method for efficiently finding good orderings.

\begin{figure}[t]
\centering
\includegraphics[width=0.5\linewidth]{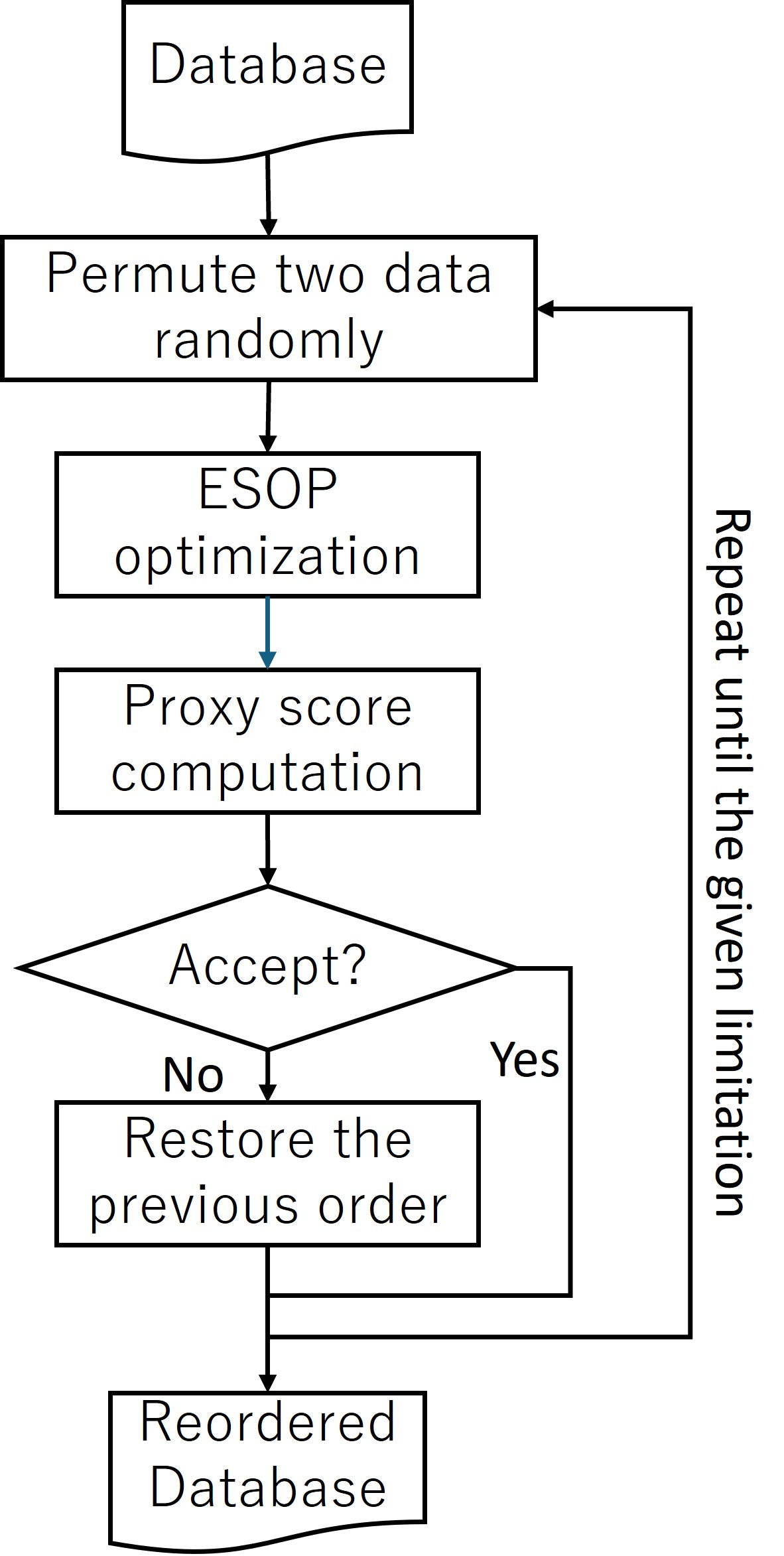}
\caption{Flow of the SA-based search in this work. Starting from an initial ordering, the process repeats data swapping, ESOP minimization, proxy evaluation, and acceptance decision to search for the best ordering.}
\label{fig:sa_flowchart}
\end{figure}

SA is a well-established optimization method; we refer the reader to~\cite{kirkpatrick1983} for details. In this paper, we use SA to iteratively modify the data ordering and search for placements with small proxy values. The basic idea is straightforward. We start with a current ordering, generate a new candidate by swapping the data at two randomly chosen addresses, perform ESOP minimization on the candidate, and compute the proxy. If the candidate is better than the current ordering, it is accepted. Moreover, even if the candidate is worse, it is accepted with a certain probability in the early stages of the search, preventing the search from being trapped in locally optimal solutions.
This allows global exploration in the first half and focused exploitation of promising regions in the second half. Fig.~\ref{fig:sa_flowchart} shows the flow of the SA-based search in this work. 

It is important to note that the novelty of this work does not lie in SA itself. The new contribution is applying SA to the problem of reducing QROM-type oracle circuits by treating the database ordering as a search variable and using the ESOP-based proxy for evaluation. Thus, this paper does not treat the algorithmic details of SA as its main subject, but rather adopts SA as a practical optimizer for ordering search.

\section{Experiments}\label{sec:experiments}

\subsection{Experimental Setup}
Experiments were conducted on a virtualized environment (WSL2, Ubuntu 24.04) running on a Windows 11 laptop equipped with an Intel Core i7-1370P CPU and 32\,GB of memory (16\,GB allocated to the virtual environment). The software stack used was Python 3.11.10 and Qiskit 2.2.2. For SA, we used simanneal\footnote{\url{https://github.com/perrygeo/simanneal}} library with default temperature settings, but with a search budget limited to 1{,}000 steps as default. For ESOP minimization, we used EXORCISM-4\footnote{\url{https://github.com/boschmitt/exorcism}}~\cite{mishchenko2001}. It was run deterministically in our experiments, always returning the same ESOP representation for the same input. Therefore, variations in the proxy are attributable solely to differences in the ordering.

In this section, we evaluate the effectiveness of the proposed method from five research questions (RQs). First, we investigate the extent to which database ordering affects ESOP complexity through exhaustive search on small-scale problems (RQ1). Second, we verify whether SA can actually find such low-cost orderings (RQ2). Third, we evaluate how well the proxy metric used during the search approximates the actual quantum circuit cost after compilation, particularly the $T$-count (RQ3). Fourth, we examine whether SA remains more effective than random search as the problem size increases (RQ4), and finally, how improvement continues as the evaluation budget grows (RQ5).

Databases were randomly generated, with each data item having a bit length of 6. For small-scale evaluation, we used 15 instances with $N=8$. For each instance, all $8!=40{,}320$ permutations were evaluated, and the circuit size after ESOP minimization was computed for each ordering. This yields the minimum, maximum, and average values for each instance. The SA evaluation budget was set to 1{,}000 steps, and the comparison baseline of random search also used 1{,}000 candidate evaluations.

For $N=16, 32, 64, 128$, the initial ordering was used as the baseline for evaluation. Since the databases were randomly generated, the initial ordering has no special significance. We use the baseline with the expectation that its proxy value is neither extremely good nor extremely poor.

\subsection{RQ1: Does database ordering affect ESOP complexity?, RQ2: Can SA find near-optimal orderings? ($N=8$)}
\begin{figure}[t]
\centering
\includegraphics[width=\linewidth]{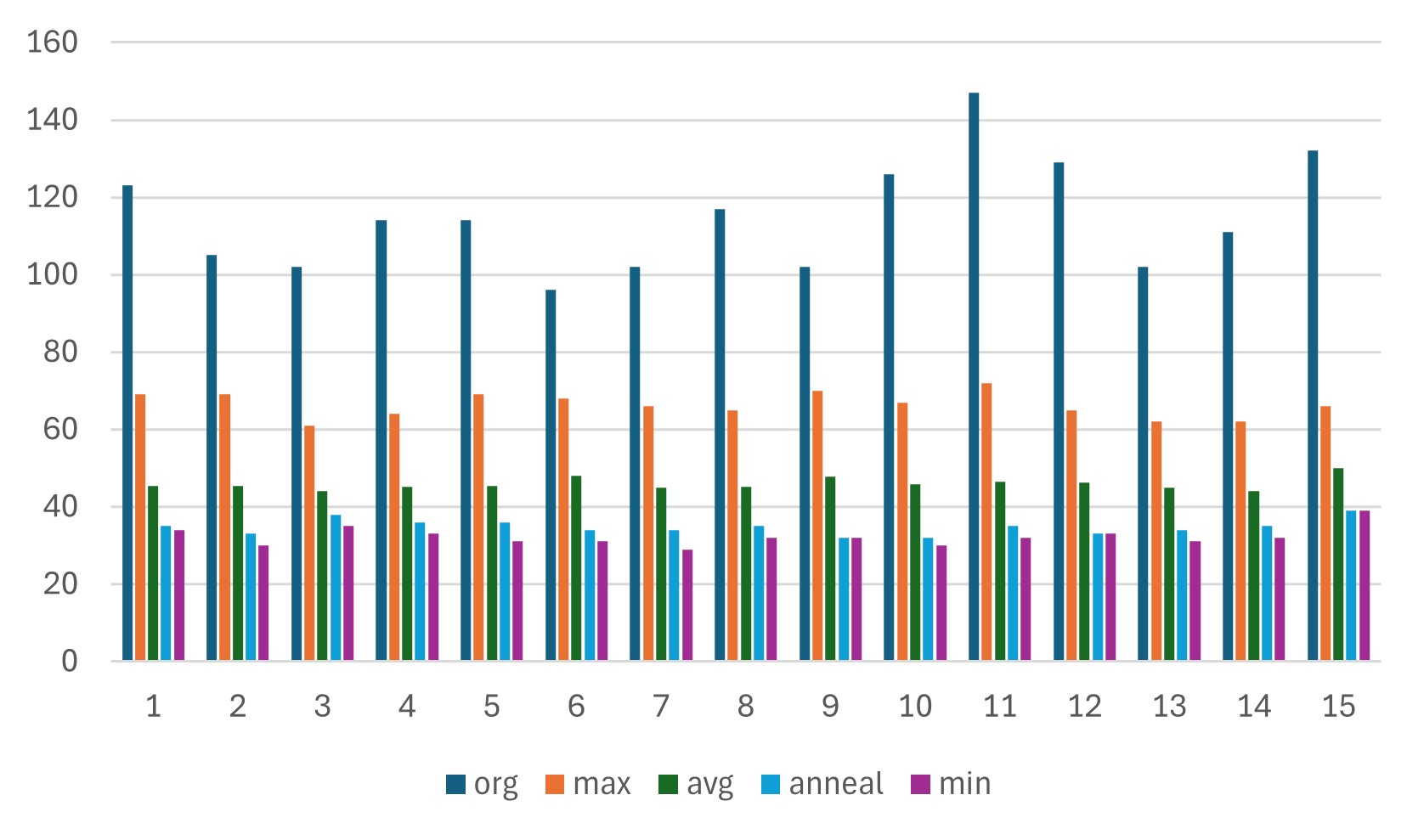}
\caption{Comparison results for 15 instances with $N=8$. org denotes the circuit size before ESOP minimization for the original ordering; max/avg/min denote the maximum, average, and minimum circuit sizes obtained after ESOP minimization over all 40{,}320 permutations; anneal denotes the circuit size after ESOP minimization for the ordering found by SA.}
\label{fig:rq12}
\end{figure}

Fig.~\ref{fig:rq12} compares, for 15 instances with $N=8$, the circuit size before ESOP minimization (org), the maximum (max), average (avg), and minimum (min) circuit sizes obtained after ESOP minimization over all 40{,}320 permutations, and the circuit size obtained after ESOP minimization for the ordering found by SA (anneal). The exhaustive search required approximately one hour. SA completed less than ten seconds.

First, the relationship between org and max shows that ESOP minimization itself is effective for circuit reduction. Applying ESOP minimization even with the original ordering significantly reduces the circuit size compared with the pre-optimization value. On the other hand, as the figure shows, the post-ESOP-minimization values vary substantially depending on the ordering. The difference between max and min is clear for each instance, and the exhaustive-search-based statistics yield a median max/min ratio of 2.097, with an interquartile range (IQR) of 2.031--2.267. This means that, even when the same ESOP minimizer is applied to the same data set, the circuit size can vary by approximately a factor of two due to differences in address assignment.

Moreover, the minimum is approximately 70\% of the average, indicating that simply choosing a random ordering is insufficient, and there is value in explicitly searching for low-cost orderings.

About the SA results, it achieves values close to min across all instances. For some instances, it matches the minimum obtained by exhaustive search; for the others, the gap is small. Therefore, for small-scale problems, SA can reliably discover the best ordering or one very close to it. This indicates that SA effectively finds the small circuits in the permutation space.

An important point is that SA achieves this level without exhaustive search in a short time. That is, the exhaustive search confirms that orderings offer substantial room for improvement, and SA demonstrates that this improvement can be realized in practice within a feasible number of evaluations.

\conclusionbox{Exhaustive search for $N=8$ confirms that data ordering can change circuit size by up to approximately a factor of two. Furthermore, SA achieves solutions close to the minimum across all instances. Database reordering is therefore an effective optimization axis, and SA is a promising search method for small-scale problems.}

\subsection{RQ3: Does the proxy metric predict T-count reduction? ($N=8$)}

Since generating and compiling a quantum circuit at every search step is expensive, this work uses an ESOP-based proxy to evaluate candidate orderings. However, if the proxy does not sufficiently reflect the actual quantum circuit cost, the orderings found by the search cannot be guaranteed to be truly advantageous. We therefore performed compilation for the $N=8$ instances and examined the relationship between the proxy ratio and the $T$-count ratio.

\begin{figure}[t]
\centering
\includegraphics[width=0.9\linewidth]{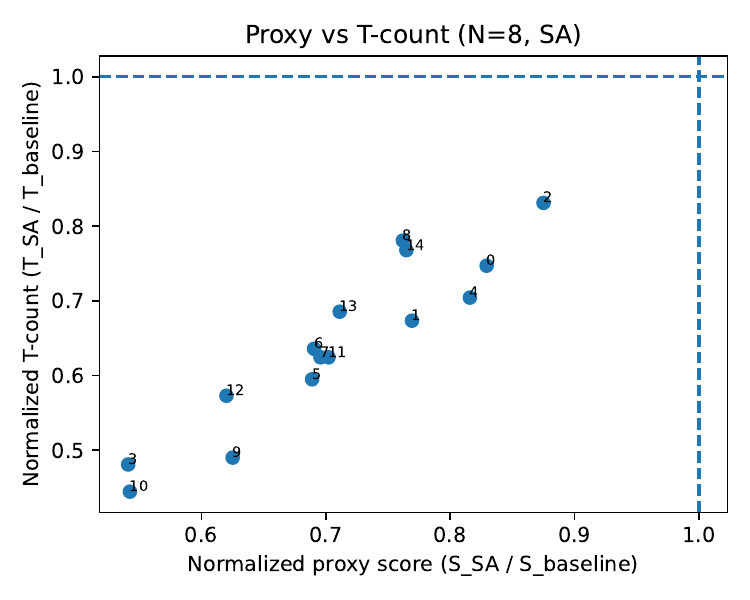}
\caption{Relationship between the proxy ratio and the $T$-count ratio for the orderings obtained by SA. Each number is the instance number from 0 to 14. A positive correlation is observed, indicating that proxy minimization is consistent with compiled circuit reduction.}
\label{fig:rq3}
\end{figure}

Fig.~\ref{fig:rq3} shows a scatter plot of the SA/baseline proxy ratio versus the SA/baseline $T$-count ratio for each instance. Here, SA refers to the ordering obtained by SA, and baseline refers to the initial ordering. Evaluating the correlation coefficient yields a Spearman correlation of $\rho=0.91$, indicating a strong positive correlation. This is consistent with the fact that, under Qiskit's default MCX decomposition method~\cite{Khattar_2025}, an $N$-controlled MCX gate is decomposed into $8n-12$ $T$ gates.

This result shows that, although the proxy does not perfectly reproduce the compiled circuit cost, it is sufficiently effective as a surrogate metric for the search. The imperfect correlation can be attributed to transpiler optimizations, local gate cancellations, and other factors.

\conclusionbox{Proxy minimization has a strong positive correlation with compiled $T$-count reduction.}

\subsection{RQ4: Is SA effective for large databases? ($N=16, 32, 64, 128$)}

\begin{figure}[t]
\centering
\includegraphics[width=\linewidth]{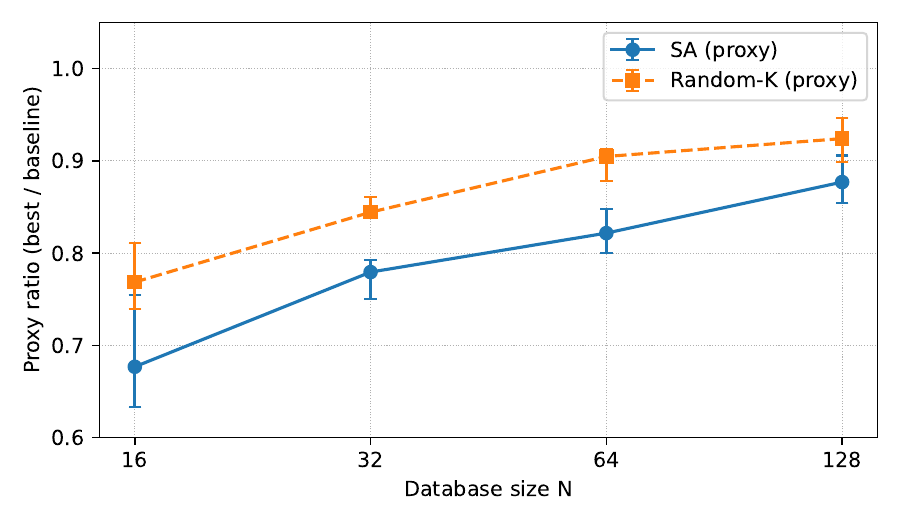}
\caption{Comparison between SA and random search (Random-K) for large instances. Each point represents the median over 15 instances; error bars show the IQR. SA yields smaller values than Random-K for all $N$.}
\label{fig:rq4}
\end{figure}

While SA can find near-optimal orderings for small-scale problems $N=8$, the practical question is whether it remains effective in large-scale settings where exhaustive search is infeasible. Fig.~\ref{fig:rq4} shows the comparison between SA and random search (Random-K) for $N=16, 32, 64, 128$ under the same evaluation budget. The vertical axis represents the ratio of the best score to the baseline, where smaller values are better. Each point represents the median over 15 instances, and error bars indicate the IQR.

For all sizes, the SA curve lies below that of Random-K. This means that, even with the same evaluation budget, SA finds better orderings. Therefore, the effect of ordering search is not limited to small-scale examples but extends to larger databases.

On the other hand, as $N$ increases, the ratios for both methods tend to approach 1. This is because the search space expands rapidly, and the amount of improvement obtained per fixed evaluation budget becomes relatively smaller. Nevertheless, the fact that SA consistently outperforms random search demonstrates that exploiting the local structure of the permutation space is effective.

\conclusionbox{For $N=16$ to $128$, SA consistently discovers better orderings than random search under the same evaluation budget.}

\subsection{RQ5: Does performance improve with more evaluation steps? ($N=64$)}
\begin{figure}[t]
\centering
\includegraphics[width=\linewidth]{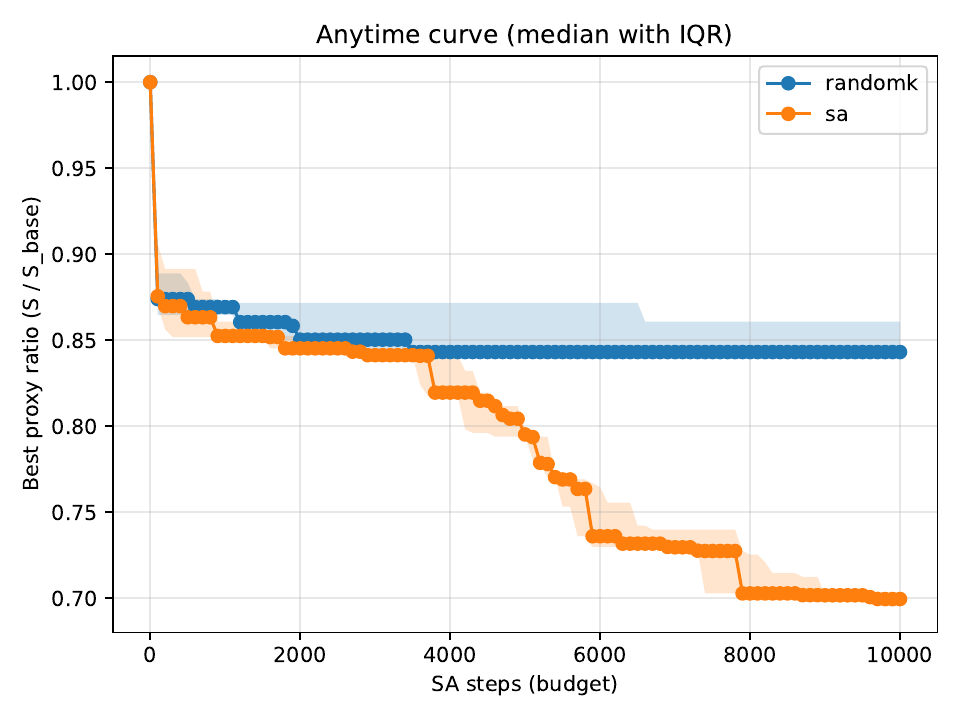}
\caption{Anytime curve for $N=64$. Random-K shows almost no improvement beyond approximately 3{,}000 steps, whereas SA continues to improve.}
\label{fig:rq6}
\end{figure}

Finally, to investigate how long it is practical to continue SA, we increased the evaluation budget for $N=64$ and recorded the score change. Fig.~\ref{fig:rq6} shows the resulting anytime curve. The horizontal axis represents the number of evaluations, and the vertical axis represents the best proxy ratio relative to the baseline (original ordering); lower values indicate greater improvement.

Random search (Random-K) improves to some extent immediately after the search begins but plateaus at approximately 3{,}000 steps. Beyond that point, increasing the evaluation budget yields almost no change in the score. In contrast, SA continues to improve in stages beyond 3{,}000 steps, demonstrating that continued search can reach lower-cost orderings. In other words, SA not only finds good candidates early but can also further explore low-cost regions in the later stages of the search.

This result demonstrates the utility of SA. In practical use, it may also be effective to set a stopping criterion based on whether the score has not been updated for a given period, in addition to a fixed step count.

\conclusionbox{Random search stops improving beyond 3{,}000 steps, whereas SA continues to improve. Running SA for as long as time and evaluation budget permit yields better solutions.}

\section{Discussion and Limitations}\label{sec:limitations}
\subsection{Validity and Limitations of the Proxy Metric}
As shown in RQ3, the proxy proposed in this paper has a strong correlation with compiled $T$-count reduction. This result is consistent with the fact that the number of cubes and specified literals in the ESOP representation correspond to a certain degree with the final MCX decomposition cost. However, the correlation is not perfect. This is because the final $T$-count depends not only on ESOP complexity but also on the MCX decomposition method, auxiliary qubit availability, transpiler optimizations, and local gate cancellations.

For future hardware-specific optimization, there is room to replace the current simple proxy with a hardware-aware proxy that incorporates, for example, per-control-count weighting or connectivity constraints. Additionally, since CX gates (control count 1) are Clifford gates that do not consume $T$ gates, the current proxy may overestimate the $T$-count. Introducing per-control-count weighting could further improve the accuracy of the proxy.

RQ3 demonstrated that the proxy correlates under $N=8$. However, as $N$ increases, the structure of ESOPs and the behavior of MCX decomposition may change, and it has not been verified whether a comparable correlation is maintained. This verification remains future work.

\subsection{Role of Simulated Annealing (SA) and Search Budget}
In RQ1/RQ2, SA was confirmed to reach the optimal ordering across all instances for $N=8$. In RQ4, SA was shown to consistently outperform Random-K in large-scale settings. Based on these results, we characterize SA in this work as follows: for small-scale problems, it functions as an optimal-solution finder, while for large-scale problems, it serves as a search method that provides high-quality approximate solutions.

However, whether SA can obtain good approximate solutions for even larger problems is uncertain. In practice, settings such as the number of search and temperature steps become important.

\subsection{Applicable Problem Class}
The premise of this work is that the ordering within the database is meaningless and can be freely changed. When this condition is satisfied, changing the mapping between addresses and data preserves the meaning of the Grover search.

On the other hand, the method is not applicable to problems where the addresses themselves have meaning. For instance, if the ordering relation or positional information of addresses is part of the specification, simple reordering is not permitted. When applying this method, it is necessary to verify that the prerequisite is satisfied.

\section{Conclusion \& Future Works}\label{sec:conclusion}

In this paper, we proposed the Database Reordering method, which treats the ordering of the database itself as an optimization variable, targeting QROM circuits within Grover oracles. In this method, ESOP minimization is applied for each ordering, and candidate orderings are evaluated using a proxy metric that captures the complexity of the resulting ESOP. Furthermore, simulated annealing (SA) is combined to search the vast permutation space within practical time.

In our experiments, exhaustive search over 15 instances with $N=8$ confirmed that data ordering significantly affects ESOP complexity. Applying SA to the same setting yielded solutions close to the minimum across all instances, demonstrating that SA is an effective search method for small-scale problems.
Furthermore, it is confirmed that there is a strong positive correlation between the proxy ratio and the $T$-count ratio, demonstrating that the proxy metric is a valid surrogate for the search. In addition, for large-scale random instances with $N$ ranging from 16 to 128, SA consistently discovered better orderings than random search.

These results demonstrate that database ordering is an effective optimization axis for QROM-type oracle circuits, and that the combination of an ESOP-based proxy and SA can practically exploit this degree of freedom. Moreover, since our method adjusts data placement as a preprocessing step rather than replacing the QROM circuit construction itself, it is compatible with existing QROM optimization and Grover oracle optimization techniques.

Future work includes, first, designing proxy metrics that more directly reflect hardware constraints and MCX decomposition methods. Second, it is important to compare SA with other search methods, such as genetic algorithms and beam search, to investigate search strategies better suited to the permutation space. Third, we plan to examine whether the proposed method can be applied to larger-scale, more practical datasets.

\bibliographystyle{IEEEtran}
\bibliography{main}
\end{document}